\documentclass[10pt,twocolumn,letterpaper]{article}

\usepackage{cvpr}              
\definecolor{cvprblue}{rgb}{0.21,0.49,0.74}
\usepackage[pagebackref,breaklinks,colorlinks,allcolors=cvprblue]{hyperref}
\usepackage{multirow}
\usepackage{graphicx}
\usepackage{float}
\usepackage{cuted}

\def\paperID{*****} 
\def\confName{CVPR}
\def\confYear{2026}

\title{2ndMatch: Finetuning Pruned Diffusion Models via Second-Order Jacobian Matching}

\author{%
Caleb Zheng \\
University of Washington\\
{\tt\small zheng94@uw.edu}
\and Eli Shlizerman \\
University of Washington\\
{\tt\small shlizee@uw.edu}
}

\begin{document}
\maketitle

\begin{strip}
\centering
\includegraphics[width=1.0\textwidth]{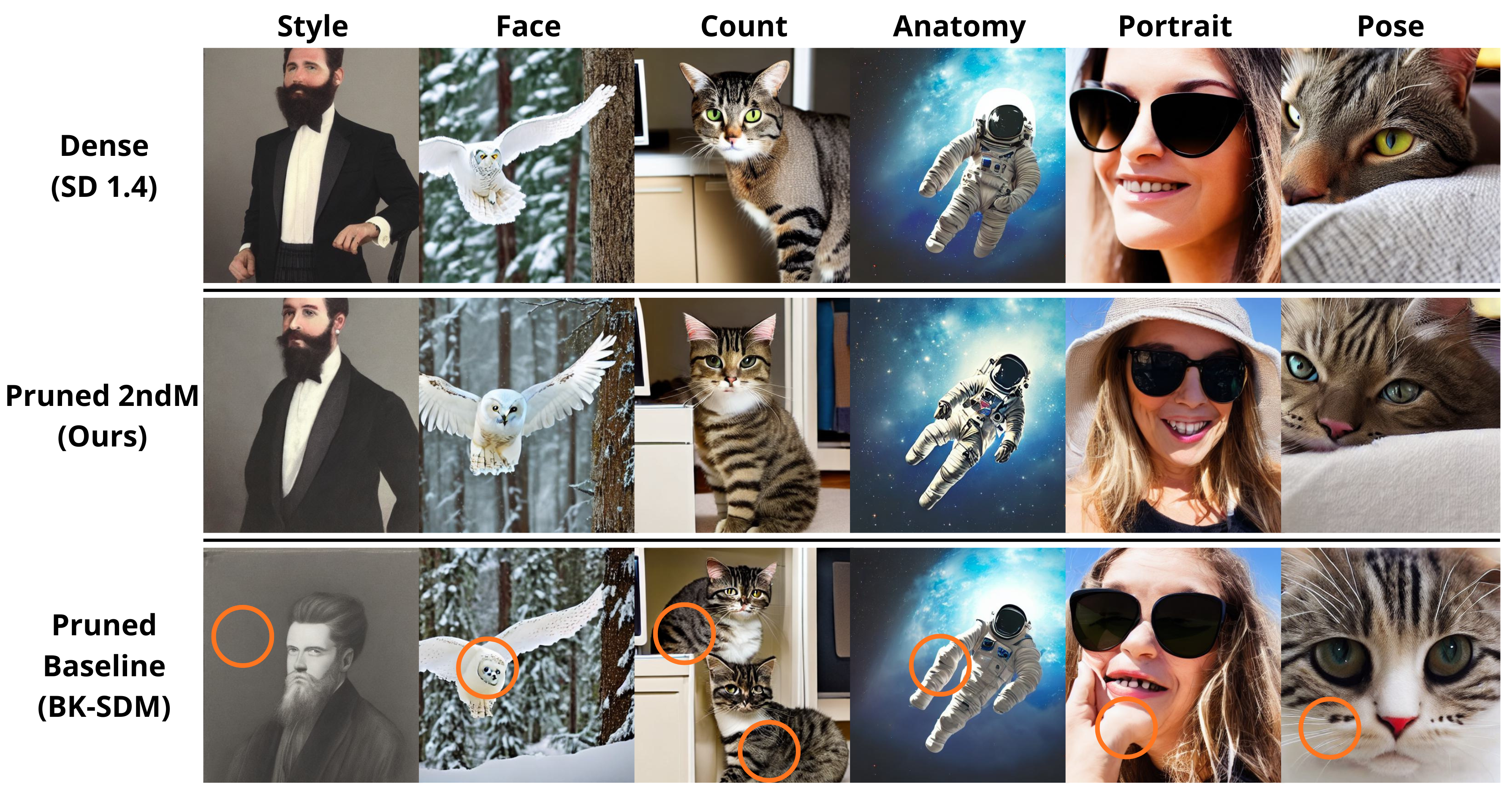}
\captionof{figure}{Examples of improvement in preserving fidelity and consistency of output images compared to baseline (BK-SDM) on pruning Stable Diffusion (SD) 1.4, Dense teacher, to 33\%.}
\end{strip}

\begin{abstract}
Diffusion models achieve remarkable performance across diverse generative tasks in computer vision, but their high computational cost remains a major barrier to deployment. Model pruning offers a promising way to reduce inference cost and enable lightweight models. However, pruning leads to quality drop due to reduced capacity. A key limitation of existing pruning approaches is that pruned models are finetuned using the same objective as the dense model (denoising score matching). Since the dense model is accessible during finetuning, it warrants a more effective approach for knowledge transfer from the dense to the pruned model. Motivated by this, we propose \textbf{2ndMatch} (\textbf{2ndM}), a general-purpose finetuning framework that introduces a \textbf{2nd}-order Jacobian ($J^{\top} J$) \textbf{M}atching loss  inspired by Finite-Time Lyapunov Exponents. \textbf{2ndM} teaches the pruned model to mimic the sensitivity of the dense teacher, i.e., how to respond to small perturbations over time, through scalable random projections. The framework is architecture-agnostic and applies to both U-Net- and Transformer-based diffusion models. Experiments on CIFAR-10, CelebA, LSUN, ImageNet, and MSCOCO demonstrate that \textbf{2ndM} reduces the performance gap between pruned and dense models, substantially improving output quality.
\end{abstract}

\begin{figure*}[ht]
    \centering
    \includegraphics[width=0.9\linewidth]{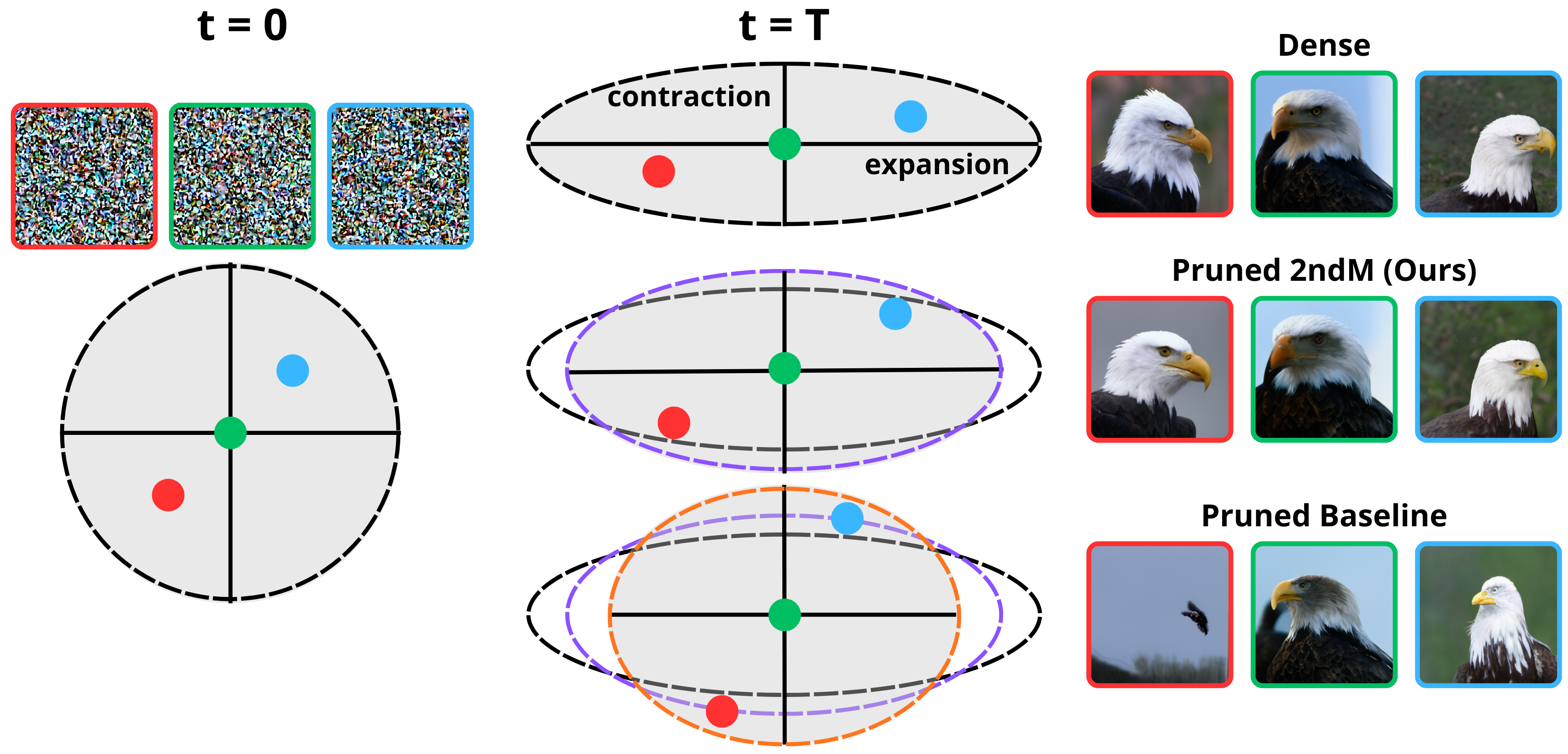}
    \caption{Sketch illustration of the difference between pruned Baseline (orange) and our \textbf{2ndM} (purple). Both Baseline and 2ndM match the dense teacher (black) on training samples (green). While 2ndM constraints the expansion/contraction behavior of nearby points (red and blue), the Baseline does not have such constraints. Due to these, the Baseline accumulates drifts along the denoising trajectory, leading to deviations at $t = T$. In contrast, 2ndM preserves stable, teacher-aligned trajectories, resulting in higher-fidelity generation.}
    \label{fig:demo}
\end{figure*}
\section{Introduction}
\label{introduction}
Diffusion models have demonstrated success across a wide range of generative tasks~\citep{ho2020denoising, saharia2022photorealistic, parmar2023zero, yang2023diffusion, li2022srdiff, wang2025videoscene}. This success, however, comes at a significant computational cost: the sampling process typically involves hundreds of iterative denoising steps, each involving a full forward pass through a large backbone architecture such as U-Net~\citep{rombach2022high} or Transformer~\citep{peebles2023scalable, bao2023all}. This computational burden presents a major challenge for real-time application and deployment in resource-constrained environments.

To mitigate the computational cost, recent efforts have improved the efficiency of diffusion models mainly addressing two complementary directions: i) reducing the number of denoising steps~\citep{song2020denoising, nichol2021improved, ma2025diffusion}, and ii) compressing the backbone architectures to lower the computational cost per step~\citep{shang2023post, li2023q, kim2024bk}. Within the compression paradigm, pruning has emerged as a promising strategy, as it directly reduces both the parameter count and the Multiply-Accumulate Operations (MACs). A growing body of work is proposing pruning techniques specifically tailored to diffusion models, incorporating structured sparsity, block-wise importance scoring, trajectory-aware objectives, and subnetwork searching~\citep{fang2023structural, castells2024ld, zhang2024effortless, fang2025tinyfusion, zhu2024dip}. 

Despite these advantages, pruned diffusion models still exhibit noticeable quality degradation compared to their dense counterparts~\citep{zhu2024dip}. A key reason lies in finetuning: most methods reuse the same objective designed for the dense model, typically denoising score matching (DSM) - which supervises the model to predict the added noise at each timestep. While effective for full-capacity models, DSM alone is insufficient for pruned models, since their reduced capacity leads to sensitivity to perturbations deviating from that of the dense teacher, as shown in Figure.~\ref{fig:demo}.

In this work, we revisit the finetuning objective in the context of pruning. Specifically, since the dense model remains accessible during finetuning, it offers an opportunity for richer supervision. Prior works have explored output- and feature-level knowledge distillation, aligning predictions or intermediate activations between the pruned and the dense models~\citep{kim2024bk}. While such strategies offer quality improvements, they still treat the dense model primarily as a source of soft labels and overlook its sensitivity - how the score responds to small perturbations across timesteps.

A natural alternative objective is \emph{Jacobian matching}, which aligns the gradients of the score function (i.e., the derivative of the predicted noise w.r.t. the input) between the pruned and the dense model. Direct (first-order) Jacobian supervision is equivalent to providing guidance to the neighborhood around each input and has shown effective in image classification~\citep{czarnecki2017sobolev, srinivas2018knowledge}. However, it is considerably less effective for diffusion models. Under the diffusion framework, inputs are already perturbed by noise and supervision on these noise-corrupted inputs is mathematically equivalent to first-order Jacobian matching. This equivalence diminishes the value of explicitly matching Jacobians, while incurring additional computational overhead. Beyond the efficiency concern, diffusion models operate as discrete-time dynamical systems, where sample quality depends not only on local (one-step) sensitivity but on how perturbations propagate over time. First-order Jacobians capture only instantaneous responses at individual timesteps and fail to reflect the long-range temporal behavior that governs the stability and fidelity of the denoising trajectory.

To address these limitations, we propose \textbf{2ndMatch} (\textbf{2ndM}), a general-purpose finetuning objective for pruned diffusion models. Instead of directly aligning first-order derivatives, \textbf{2ndM} aligns the pruned model with its dense teacher in terms of \emph{temporal} behavior—specifically, how the score (denoising field) responds over time to small input perturbations. This perspective is inspired by \emph{Finite-Time Lyapunov Exponents} (FTLE), which quantify how infinitesimal perturbations \emph{expand} or \emph{contract} over finite horizons, see Figure.~\ref{fig:demo}. FTLE provides a principled lens on sampling stability and sensitivity. However, the computation of FTLE for high-dimensional diffusion models is computationally prohibitive, as it requires evaluating and tracking full Jacobians across multiple timesteps. To maintain scalability, 2ndM approximates this process via random projection and a step-wise matching, estimating sensitivity via directional quadratic forms $v^{\top} J^{\top} J v$, efficiently estimating how perturbations evolve through the denoising process.

Our contributions are summarized as follows:
\begin{itemize}
    \item We identify key limitations in existing finetuning approaches for pruned diffusion models, including those based on output-level knowledge distillation or first-order Jacobian matching. These do not guide the pruned model to preserve the dense model’s sensitivity to perturbations.
    \item We propose \textbf{2ndMatch} (\textbf{2ndM}), an architecture- and pruning-method-agnostic finetuning framework inspired by FTLE that matches the \emph{second-order} Jacobian metric $J^\top J$ (not Hessians) to align the dense teacher and pruned student’s temporal sensitivity to small perturbations, using random projections and a step-wise matching to estimate $v^\top J^\top J v$ without multi-step Jacobian tracking.
    \item Our experiments demonstrate that \textbf{2ndMatch} consistently narrows the quality gap between pruned and dense models, and outperforms existing pruning approaches across both U-Net-based and Transformer-based diffusion models, evaluated on CIFAR-10, CelebA, LSUN, ImageNet, and MSCOCO datasets.
\end{itemize}
\section{Related Work}
\paragraph{Efficient Diffusion Models.}
Diffusion models require multiple forward passes through large backbones and thus are computationally expensive~\citep{rombach2022high,peebles2023scalable, bao2023all}. One line of work accelerates inference by reducing the number of denoising steps through schedule optimization~\citep{watson2021learning, ye2025schedule, ding2025rass}, ODE/SDE-based solvers~\citep{song2020denoising, song2020score, lu2022dpm, zhang2022fast, karras2022elucidating} or distillation-based samplers~\citep{salimans2022progressive, luhman2021knowledge, meng2023distillation, zheng2024diffusion, cai2025shortcutting, ma2025diffusion}. While effective, these approaches do not necessarily reduce storage or memory overhead during inference, limiting their practicality in resource-constrained settings. 

Another direction targets per-step efficiency by alternating between models~\citep{liu2023oms, pan2024t, zhang2024accelerating}, caching intermediate features~\citep{agarwal2024approximate, ma2024deepcache, wimbauer2024cache, liu2025timestep}, or compressing the backbone through quantization~\citep{shang2023post, li2023q, zhao2025pioneering, chen2025q} and pruning~\citep{fang2023structural}. Among these strategies, pruning has attracted growing attention. Diff-Pruning applied structured pruning, removing entire groups of weights for hardware-friendly acceleration~\citep{fang2023structural}, while subsequent methods estimate the importance of network components using heuristics or auxiliary networks~\citep{castells2024ld, zhang2024laptop, zhu2024dip}. EcoDiff introduces a differentiable pruning mask trained end-to-end across the sampling trajectory, capturing cumulative temporal effects but not explicitly regularizing the score function’s sensitivity~\citep{zhang2024effortless}.

While these methods improve computational efficiency, a quality gap between pruned and dense models persists, largely due to reusing the original objective, denoising score matching (DSM), during finetuning. Some works extend DSM with knowledge distillation~\citep{hinton2015distilling}, aligning the outputs or intermediate features of the pruned and dense models~\citep{kim2024bk, ganjdanesh2024not, shirkavand2025efficient}. While these approaches were able to reduce the quality gap, they do not fully leverage the dense model’s capacity. We argue that more effective use of the dense model, beyond soft labels, is possible, particularly in guiding the sensitivity of the pruned model. This motivates revisiting score matching and exploring derivative-based supervision to provide richer finetuning signals.
\vspace{-0.45cm}
\paragraph{Score Matching and Jacobian Matching.}
Score-based generative models aim to approximate the gradient of the data log-density, i.e., score, which in diffusion models is typically learned through DSM~\citep{hyvarinen2005estimation}. While DSM provides strong supervision at each timestep, it does not constrain how the score evolves over time, leaving temporal dynamics under regularized. In pruned models with reduced capacity such under regularization becomes more pronounced.

Prior work has investigated derivative-based alignment, demonstrating that incorporating higher-order derivatives of the teacher can improve student learning~\citep{czarnecki2017sobolev}. First-order Jacobian matching is mathematically equivalent to enforcing output consistency under perturbed inputs and has been proven effective for image classification~\citep{srinivas2018knowledge}. For diffusion models, however, it does not transfer well due to their stochastic and iterative generation process. In particular, the stochasticity of the diffusion process makes first-order Jacobian matching effectively equivalent to output-level knowledge distillation. Moreover, since diffusion models evolve sequentially, their performance depends not only on one-step input-output Jacobian but also on how perturbations propagate and accumulate across timesteps. First-order Jacobians do not capture such behavior. This limitation motivates rethinking finetuning objectives through the lens of dynamical systems theory, where long-range sensitivity and stability can be explicitly regularized.
\vspace{-0.4cm}
\paragraph{Finite-Time Lyapunov Exponents.}
To capture how perturbations propagate over time, dynamical systems theory provides valuable tools for finetuning objective design. 

Diffusion models can be viewed as discrete-time dynamical systems, where each denoising step updates samples based on the predicted noise, enabling sensitivity analyses across timesteps. A classical measure of such sensitivity is the Lyapunov Exponent (LE), which quantifies the expansion/contraction rate of nearby trajectories~\citep{ruelle1979ergodic, oseledets2008oseledets}. While LE has been used to study stability in deep learning~\citep{vogt2024lyapunov, zheng2025hyperpruning}, the assumption of LE of time-invariance and its reliance on analytical expression of the dynamics make LE unsuitable for time-varying, high-dimensional systems such as diffusion models.

The \emph{Finite-Time Lyapunov Exponent} (FTLE) addresses these limitations by measuring the local expansion/contraction rate over a finite horizon via second-order Jacobians. FTLE has been broadly applied in domains such as fluid dynamics and biomedical modeling~\citep{shadden2005definition, brunton2010fast, lipinski2010ridge, haller2015lagrangian}. 

Its use in high-dimensional model remains computationally expensive, as it requires evaluating full Jacobians over space and time. Recent advances in machine learning have introduced scalable ways to estimate Jacobian related quantities without computing the full Jacobians. For instance, SSM estimates score gradients via random-direction projection~\citep{song2020sliced, song2019generative}. Inspired by this idea, we develop a \emph{second-order Jacobian matching loss} rooted in FTLE theory. By projecting Jacobians into low-dimensional subspaces and aligning the sensitivity patterns between the dense and the pruned model, our method improves the sample quality of pruned model without requiring full Jacobian computation.
\vspace{-0.5cm}
\section{Background}
\vspace{-0.1cm}
\label{bakcground}
In this section, we review Finite-time Lyapunov Exponent (FTLE) theory and fundamental concepts in score-based generative models.
\vspace{-0.3cm}
\paragraph{Finite-time Lyapunov Exponent.}
The FTLE characterizes the expansion or contraction rate of infinitesimally close trajectories in time-dependent dynamical systems, governed by $\frac{dx(t)}{dt} = f(x(t), t)$, where $f(x(t), t): \mathbb{R}^n \times \mathbb{R} \rightarrow \mathbb{R}^n$ describes the system dynamics,  $x(t)\in \mathbb{R}^n$ is the state variable, and $t \in \mathbb{R}$ denotes time. Given an initial condition $x_0$, the trajectory over the interval $[t_0, t_0 + t_1]$, with $t_1 \in \mathbb{R}^+$, is obtained by numerically integrating the dynamics. We denote the flow map by $\Phi_{t_0}^{t_0 + t_1}: \mathbb{R}^n \rightarrow \mathbb{R}^n$, which maps the initial state $x_0$ to the final state $x_1$
\begin{align}
    \Phi_{t_0}^{t_0 + t_1}: x_0 \rightarrow x_0 + \int_{t_0}^{t_0 + t_1}f(x(\tau), \tau) d \tau.
\end{align}
Next, considering a nearby point 
$x^{\prime}_0 = x_0 + v_0$, where $v_0 \in \mathbb{R}^n$ is an infinitesimal perturbation, we can write the value of the perturbation at $t_1$, $v_1 = v(t_1)$, according to a first-order Taylor expansion
\begin{align}
    v_1 & = \nabla_{x_0} \Phi_{t_0}^{t_0 + t_1}(x_0) v_0 + \mathcal{O}(\| v_0\|^2).
\end{align} 
Assuming that the higher-order term is negligible for infinitesimal $v_0$, the magnitude of $\|v_1\| $ representing the distance of the perturbation after $t_1$ is approximated by
\begin{align}
    \|v_1\| \approx \sqrt{v_0^{\top} J^{\top} J v_0},
    \label{eq:2nd-order_jac}
\end{align}
where $J = \nabla_{x_0} \Phi_{t_0}^{t_0 + t_1}(x_0)$ is the Jacobian of the flow map and $(\cdot)^{\top}$ denotes the matrix transpose operation. The maximum amplification occurs when $v_0$ aligns with the eigenvector associated with the maximum eigenvalue $\lambda_{\max}$ of $J^{\top} J$. The resulting maximum perturbation magnitude is then $\sqrt{\lambda_{\max}}\|v_0\|$ from which the associated FTLE is defined as the average exponential stretching ratio in the direction of maximal stretching $\Lambda^{t_0 + t_1}_{t_0}(x_0) = \frac{1}{t_1}\ln \sqrt{\lambda_{\max}}$.
\vspace{-0.6cm}
\paragraph{Score-based Generative Modeling.}
Given a dataset $\{x_i\}_{i=1}^D$, we assume there exists an intractable data distribution $p(x)$ from which each data point is independently sampled. 
Score-based generative modeling aims to sample from $p(x)$ by learning the gradient of the log-probability density $\nabla_x \log p(x)$, also known as the score. A Noise Conditional Score Neural Networks (NCSN) $s(\cdot; \theta): \mathbb{R}^n \rightarrow \mathbb{R}^n$ is trained to approximate this score via the score matching loss
\begin{align}
    \frac{1}{2} \mathbb{E}_{x\sim p(x)}\Big[\|s(x; \theta) - \nabla_x \log p(x)\|^2_2\Big].
    \label{eq:score_matching}
\end{align}
Since the true score $\nabla_x \log p(x)$ is typically intractable, denoising score matching (DSM) approximates it by training on noise-perturbed data~\citep{vincent2011connection}. Specifically, for a given noise level $\sigma > 0$, noisy samples $\tilde{x}\sim q_{\sigma}(\tilde{x}|x)=\mathcal{N}(x, \sigma^2 I)$ are drawn and the objective becomes
{
\small
\begin{align}
    l(\theta; \sigma) = \frac{1}{2} \mathbb{E}_{\tilde x} \Big[\| s(\tilde{x}, \sigma;\theta) + \frac{\tilde{x} - x}{\sigma^2}\|^2_2\Big].
\end{align}
}NCSN extends this by estimating a family of score functions across multiple noise levels $\{\sigma_i\}_{i=1}^L$~\citep{song2019generative}. The unified multi-scale training objective is
\begin{align}\
    \mathcal{L}(\theta; \{\sigma_i\}_{i=1}^L) = \frac{1}{L} \sum_{i=1}^L \lambda(\sigma_i) l(\theta; \sigma_i),
\end{align}
where $\lambda(\sigma_i)$ is a weighting factor.

In practice, score-based diffusion models often simplify DSM by directly predicting the noise $\epsilon \sim \mathcal{N}(0, I)$ added in the forward process using MSE loss $\mathcal{L} = \| s(\tilde{x} , t;\theta) - \epsilon \|^2_2$, where the timestep $t$ indexes the noise level and can be mapped to corresponding noise level $\sigma_t$ through a predefined noise schedule and $\tilde{x}$ is the image corrupted by $\epsilon$ during the forward diffusion process. This formulation avoids explicit noise-level reweighting and works well empirically. 

However, this simplification comes with limitations. Although effective in aligning each timestep, it does not explicitly constrain the temporal behavior of the model or does not capture how perturbations evolve during generation. These limitations are especially prominent in pruned models with reduced capacity. To address these challenges, we explore Jacobian-based losses that incorporate dynamical sensitivity, enabling improved temporal fidelity. We introduce these techniques in the next section.
\vspace{-0.1cm}
\section{Method}
Given a pruned model $s(\cdot; \theta)$, our goal is to effectively finetune it so as to minimize the generative performance gap with its dense counterpart $s_{\mathcal{D}}(\cdot; \theta_{\mathcal{D}})$. To this end, we propose a hybrid finetuning objective that integrates three complementary components: Noise Prediction ($\mathcal{L}_{\text{NP}}$), output-level Knowledge Distillation ($\mathcal{L}_{\text{KD}}$), and 2nd-order Jacobian matching ($\mathcal{L}_{\text{2nd-Jac}}$). Each term targets a specific challenge associated with finetuning pruned models, particularly under reduced model capacity, and together, they form a unified framework for guiding pruned diffusion models toward teacher-level performance
\begin{align}
    \mathcal{L}_{\text{total}}
    = \lambda_{\text{NP}}\mathcal{L}_{\text{NP}}
    + \lambda_{\text{KD}}\mathcal{L}_{\text{KD}}
    + \lambda_{\text{Jac}}\mathcal{L}_{\text{2nd-Jac}} .
\end{align}
\vspace{-1.05cm}
\paragraph{Noise prediction.} $\mathcal{L}_{\text{NP}} = \mathbb{E}_{\tilde{x},t,\epsilon}\big[ \|s(\tilde{x}, t; \theta) - \epsilon\|_2^2 \big]$, where $\epsilon\sim\mathcal{N}(0,I)$. This is the standard Denoising Diffusion Probabilistic Models (DDPM) objective, which supervises the model to predict noise added during forward diffusion process. However, for pruned models, directly predicting noise can be particularly challenging due to their limited capacity, especially early in finetuning when model outputs significantly deviate from the target noise.
\vspace{-0.3cm}
\paragraph{Knowledge Distillation.} $\mathcal{L}_{\text{KD}} = \mathbb{E}_{\tilde{x},t}\big[\|s(\tilde{x}, t; \theta) - s_{\mathcal{D}}(\tilde{x}, t; \theta_{\mathcal{D}})\|_2^2\big]$. This term mitigates the instability that arises when low-capacity pruned models are trained on noisy supervision by providing smoother targets from the teacher model. Knowledge distillation thus guides the pruned model toward meaningful intermediate representations and accelerates convergence during early finetuning.

While $\mathcal{L}_{\text{NP}}$ and $\mathcal{L}_{\text{KD}}$ provide strong supervision, they treat each timestep in isolation and therefore fail to capture temporal dependencies across the denoising process. Consequently, it may insufficiently constrain how the denoising dynamics evolve over time, limiting the model’s ability to maintain stable during sampling, see an illustration in Figure~\ref{fig:demo}. To address this, we introduce a Jacobian-based loss that explicitly regularizes the pruned model’s sensitivity to input perturbations throughout the denoising trajectory.
\vspace{-0.5cm}
\paragraph{Second-Order Jacobian Matching.} Diffusion models progressively transform Gaussian noise into structured data through iterative denoising steps. At each step, the model outputs $s(x, t; \theta)$ define the direction and the magnitude of the update, while the Jacobian $J := \nabla_x s(x, t; \theta)$ captures the local sensitivity of the output to input perturbation. 
\begin{table*}[ht!]
\centering
\caption{Quantitative comparison of methods applied to UNet-based diffusion models on LSUN-Church/Bedroom (256×256), and ImageNet ($256\times256$). IS is omitted for LSUN datasets since it relies on ImageNet-pretrained classifier and is therefore not a reliable metric for LSUN evaluation. The \textbf{rFID} entry for DeepCache is marked as ``--'' since it was not reported in the original paper.}
\vspace{-0.2cm}
\label{tab:unet}
\centering
\begin{tabular}{l|cccc|cccccc} 
\toprule
\multicolumn{11}{c}{\textbf{LSUN-Church ($256 \times 256$)}} \\ 
\textbf{Method} & \textbf{\# Params $\downarrow$} & \textbf{MACs $\downarrow$} & \textbf{Speed $\uparrow$} & \textbf{Imgs/s $\uparrow$}  & \textbf{rFID $\downarrow$} & \textbf{sFID $\downarrow$} & \textbf{FID $\downarrow$} & \textbf{IS $\uparrow$} &  \textbf{Prec. $\uparrow$} & \textbf{Rec. $\uparrow$} \\
\midrule
DDPM & 113.7 M   & 248.7 G & $1.0 \times$ & 0.35/s & - & 3.61 & 10.58 & -&  61.01 & 39.27 \\
\midrule
Diff-Pruning & 63.2 M & 138.8 G & $1.34 \times$ & 0.47/s & 4.09 & 5.04 & 13.90 & - & 50.43 & \textbf{41.14} \\
Deep Cache & 113.7 M & 156.0 G & $1.48 \times$ & 0.47/s & - & - & 13.68 & & - & -\\
\textbf{2ndM (Ours)} & 63.2 M & 138.8 G & $1.48 \times$ & 0.47/s & \textbf{2.08} & \textbf{4.13} & \textbf{11.25} & - &\textbf{54.36} & 38.51 \\ 
\midrule
\multicolumn{11}{c}{\textbf{LSUN-Bedroom ($256 \times 256$)}} \\
\textbf{Method} & \textbf{\# Params $\downarrow$} & \textbf{MACs $\downarrow$} & \textbf{Speed $\uparrow$} & \textbf{Imgs/s $\uparrow$}  & \textbf{rFID $\downarrow$} & \textbf{sFID $\downarrow$} & \textbf{FID $\downarrow$} & \textbf{IS $\uparrow$} &  \textbf{Prec. $\uparrow$} & \textbf{Rec. $\uparrow$} \\
\midrule
DDPM   & 113.7 M   & 248.7 G & $1.0 \times$ & 0.35/s & - & 3.81 & 6.62 & - & 52.75 & 41.67 \\
\midrule
Diff-Pruning & 63.2 M & 138.8 G & $1.48 \times$ & 0.47/s & 7.62 & 5.84 & 17.90 & - & 36.41 & \textbf{39.98}\\
Deep Cache   & 113.7 M & 156.0 G & $1.48 \times$ & 0.47/s & - & - & \textbf{9.49} & -  & - & - \\
\textbf{2ndM (Ours)} & 63.2 M & 138.8 G & $1.48 \times$  & 0.47/s & \textbf{2.16} & \textbf{4.53} & 9.68 & - & \textbf{44.15} & 39.18 \\
\midrule
\multicolumn{11}{c}{\textbf{ImageNet ($256 \times 256$)}} \\
\textbf{Method} & \textbf{\# Params $\downarrow$} & \textbf{MACs $\downarrow$} & \textbf{Speed $\uparrow$} & \textbf{Imgs/s $\uparrow$}  & \textbf{rFID $\downarrow$} & \textbf{sFID $\downarrow$} & \textbf{FID $\downarrow$} & \textbf{IS $\uparrow$} &  \textbf{Prec. $\uparrow$} & \textbf{Rec. $\uparrow$} \\
\midrule
LDM-4   & 400.9 M & 99.8 G & $1.0 \times$ & 0.26/s & - & 3.52 & 3.6 & 265.46 & 73.54 & 46.17\\
\midrule
Diff-Pruning & 175.8 M & 43.2 G & $1.7 \times$ & 0.44/s & 9.28 & 6.48 & 10.23 & 131.85 & 60.63 & 47.53\\
\textbf{2ndM (Ours)} & 175.8 M & 43.2 G & $1.7 \times$ & 0.44/s & \textbf{4.11} & \textbf{4.41} & \textbf{5.68} & \textbf{168.40} & \textbf{68.56} & \textbf{48.92} \\
\bottomrule
\end{tabular}
\vspace{-0.4cm}
\end{table*}
Jacobian matching offers a natural way to align the local sensitivity of a pruned model with that of its dense counterpart by minimizing $||J - J_\mathcal{D}||_F^2$. However, in diffusion models, this first-order Jacobian matching is largely redundant with output-level knowledge distillation. Under noisy inputs, output alignment implicitly enforces first-order Jacobian matching through a Taylor expansion. Specifically, let $x' = x + \zeta$, where $\zeta \sim \mathcal{N}(0, \sigma^2 I)$ denotes small Gaussian perturbations around $x$. Then,
\begin{equation}
\begin{aligned}
    ||s(x^{\prime}) - s_{\mathcal{D}}(x^{\prime})||^2_2  
    = & ||s(x) - s_{\mathcal{D}}(x)||^2_2 \\
    & + \sigma^2 ||J - J_\mathcal{D}||^2_F + \mathcal{O}(\sigma^4).
\end{aligned}
\label{eq:taylor}
\end{equation}
As a result, explicit first-order Jacobian supervision offers little additional benefit while increasing computational cost. Moreover, first-order alignment fails to capture temporal sensitivity — how small perturbations evolve and accumulate across denoising steps. To address this, we introduce a \emph{second-order Jacobian matching} objective inspired by Finite-Time Lyapunov Exponents (FTLE), which measure the expansion or contraction rate of infinitesimal perturbations over finite horizons. Concretely, we align the second-order Jacobian, $J^{\top}J$, between pruned and dense models to match their temporal response characteristics. 
In practice, we use a \emph{step-wise matching} that replaces multi-step Jacobian propagation along the denoising trajectory with a local per-timestep sensitivity proxy (Supplementary Material Sec.A.3).

However, the whole Jacobian matrix is infeasible to form during training due to high-dimensionality. Therefore, rather than comparing full $J^{\top} J$, we project the Jacobian onto random directions $v \sim \mathcal{N} (0, I)$, yielding the directional expansion rate $||J\hat{v}||^2_2 = \hat{v}^{\top} J^{\top}  J \hat{v}$, where $\hat{v} = v / ||v||$ is normalized $v$. We define the matching loss as
\begin{align}
    \mathcal{L}_{\text{2nd-Jac}}
    = \mathbb{E}_{\tilde{x},t,v}\Big[\big(\|J \hat{v}\|_2^2 - \|J_{\mathcal{D}} \hat{v}\|_2^2\big)^2\Big].
\end{align}
This formulation measures discrepancies in directional expansion or contraction behavior between pruned and dense models. Importantly, it can be computed efficiently using Jacobian--vector products (JVPs) (without materializing full Jacobian matrices), which are supported by modern autograd systems such as PyTorch, making the approach scalable to large diffusion models.
\vspace{-0.1cm}
\section{Experiments}
\subsection{Experimental Setup}
We evaluate \textbf{2ndMatch} on both U-Net– and Transformer–based diffusion models. For U-Net models, we use pretrained checkpoints: DDPM~\citep{ho2020denoising} and Stable Diffusion (SDM-v1.4)~\citep{rombach2022high} models from Hugging Face, and the LDM-4~\citep{rombach2022high} from the official CompVis GitHub repository. For Transformer-based models, we adopt U-ViT~\citep{bao2023all}.  

We assess the generative quality using Fréchet Inception Distance (FID)~\citep{heusel2017gans}, relative FID (rFID)~\citep{zheng2024diffusion}, spatial FID (sFID)~\citep{nash2021generating}, Precision-and-Recall~\citep{kynkaanniemi2019improved},  Inception Score (IS)~\citep{salimans2016improved} and Clip score~\citep{hessel2021clipscore}. Unless otherwise specified, all metrics are computed over $50,000$ generated samples. For efficiency evaluation, we report the number of parameters, Multiply–Accumulate Operations (MACs), inference speed acceleration, and throughput (images/s), all measured on a single NVIDIA A100 (80\,GB) GPU.  
\begin{figure*}[t]
    \centering
    \includegraphics[width=1.0\linewidth]{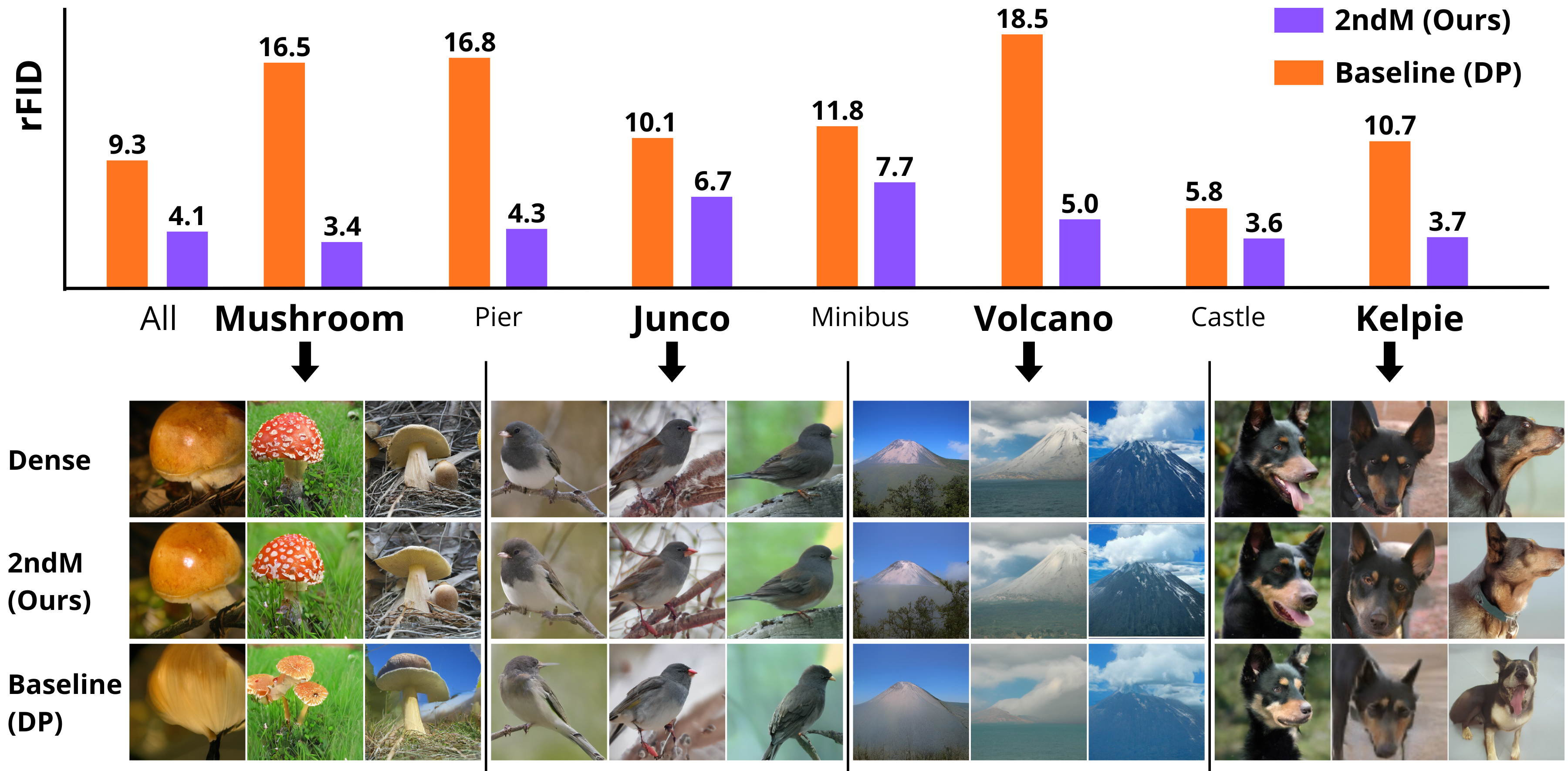}
    \vspace{-0.5cm}
    \caption{Per-class rFID and qualitative comparison between Dense, 2ndMatch, and baseline finetuning (DP) on ImageNet 256×256.}
    \label{fig:imagenet_demo}
    \vspace{-0.4cm}
\end{figure*}
We adopt Diff-Pruning~\citep{fang2023structural} as the primary pruning baseline for DDPM, LDM-4 and U-ViT, and BK-SDM~\citep{kim2024bk} for Stable Diffusion Model(SDM-v1.4). To ensure fair comparison, we follow the same training and sampling configurations as prior works~\citep{fang2023structural, kim2024bk, bao2023all}: 100-step DDIM sampling for DDPM and LDM, 25-step sampling for SDM-v1.4, and 1000-step Euler–Maruyama SDE sampling for U-ViT. Our experiments cover image resolutions ranging from $32\times32$ to $512\times512$, including CIFAR-10 ($32\times32$)~\citep{krizhevsky2009learning}, CelebA ($64\times64$)~\citep{liu2015deep}, LSUN-Church/Bedroom ($256\times256$)~\citep{yu2015lsun}, ImageNet ($256\times256$)~\citep{deng2009imagenet}, and MS COCO ($512\times512$)~\citep{lin2014microsoft}. Additional implementation and training details are provided in the Supplementary Material Sec.C.
\vspace{-0.1cm}
\subsection{Performance on U-Net-based Models}
We first evaluate 2ndMatch on U-Net–based diffusion models, including DDPM, LDM-4, and SDM v1.4, across datasets of increasing scale and complexity. Table.~\ref{tab:unet} summarizes results on LSUN-Church, LSUN-Bedroom, and ImageNet at $256\times256$ resolution. Across all datasets, 2ndMatch consistently reduces the gap between pruned and dense models in terms of FID, sFID and rFID, while preserving efficiency achieved by Diff-Pruning. 

Specifically, on the two LSUN benchmarks, 2ndMatch yields substantial improvements over Diff-Pruning at identical parameter count and MACs. On LSUN-Church, it reduces FID from $13.90$ to $11.25$, while also improving rFID from $4.09$ to $2.08$ and sFID from $5.04$ to $4.13$, indicating better alignment with both the dense reference model and the real data distribution. On LSUN-Bedroom, the gains are even larger: FID drops from 17.90 to 9.68, corresponding to a relative improvement of nearly $46\%$. While DeepCache achieves marginally better FID (9.49), it requires nearly ×2 number of parameters and incurs additional memory overhead during inference due to feature caching. On ImageNet, 2ndMatch lowers FID from \textbf{10.23} to \textbf{5.68} and improves rFID by over \textbf{55\%}, demonstrating better alignment with the dense teacher’s distribution. These results highlight that explicitly regularizing the temporal sensitivity of the score function leads to both quantitative and perceptual gains.

Figure~\ref{fig:imagenet_demo} provides qualitative comparisons on ImageNet, where samples are generated for multiple classes using both 2ndMatch (middle) and the baseline (Diff-Pruning)(bottom). The bar chart at the top shows per-class rFID improvements which consistently favor 2ndMatch. These were computed by conditioning both models on the same class and comparing $50,000$ samples. Visually, inclusion of 2ndMatch loss produces images that are more faithful to the dense reference, preserving global structure and texture details which the baseline fails to recover, particularly in complex classes such as animals or terrains.

We further apply 2ndMatch loss to prune SDM v1.4  in conjunction with BK-SDM at various model sizes.
As shown in Table \ref{tab:gen_scores}, 2ndMatch consistently improves both FID and CLIP score across Base, Small, and Tiny variants, yielding an average relative gain of \textbf{1.02} in FID while requiring no additional parameters. These improvements confirm that the proposed Jacobian based finetuning generalizes effectively to large-scale text-to-image models.
\begin{table}[h!]
\centering
\caption{Generation quality on MS-COCO ($512\times 512$, 30K samples) for BK-SDM–pruned SDM v1.4 models of different sizes (Base / Small / Tiny), with and without 2ndM finetuning.}
\vspace{-0.3cm}
\label{tab:gen_scores}
\setlength{\tabcolsep}{4pt}
\renewcommand{\arraystretch}{1.0}
\small
\begin{tabular}{lccccccc}
\toprule
\textbf{Method} & \textbf{\#Param$\downarrow$} & \textbf{MACs $\downarrow$} & \textbf{Speed $\uparrow$}& \textbf{FID$\downarrow$} & \textbf{CLIP$\uparrow$} \\
\midrule
SDM-v1.4 & 1.04B & 338.7G & 1.0 x& 13.05 & 0.2958 \\  \midrule
Base & 0.76B & 223.8 G & 1.48 x & 15.76 &  0.2878 \\
Base+2ndM & 0.76B & 223.8 G & 1.48 x & \textbf{13.84} &  \textbf{0.2900} \\ 
\midrule
Small & 0.66B & 217.7 G & 1.62 x & 16.98 &  0.2677 \\
Small+2ndM & 0.66B & 217.7 G & 1.62 x & \textbf{16.17} & \textbf{0.2682} \\  \midrule
Tiny & 0.50B & 205.0 G & 1.77 x &  17.12 &  \textbf{0.2653} \\
Tiny+2ndM & 0.50B & 205.0 G & 1.77 x& \textbf{16.68} & 0.2636\\
\bottomrule
\end{tabular}
\vspace{-0.3cm}
\end{table}
\subsection{Performance on Transformer-based Models}
We further evaluate 2ndMatch on Transformer-based diffusion models using U-ViT~\citep{bao2023all} and compare performance against state-of-the-art compression methods including Diff-Pruning~\citep{fang2023structural} and SparseDM~\citep{wang2024sparsedm}. Table \ref{tab:transformer} reports the results on CIFAR-10 ($32\times32$) and CelebA ($64\times64$).
\begin{table}[ht]
\centering
\setlength{\tabcolsep}{3pt}
\renewcommand{\arraystretch}{1.0}
\caption{Quantitative comparison on Transformer-based diffusion models (U-ViT) for CIFAR-10 ($32\times32$) and CelebA ($64\times64$).}
\vspace{-0.3cm}
\label{tab:transformer}
\begin{tabular}{lcccc}
\toprule
\multicolumn{5}{c}{\textbf{CIFAR-10 ($32\times32$)}} \\
\textbf{Method} & \textbf{\# Param $\downarrow$} & \textbf{MACs$\downarrow$} & \textbf{Speed $\uparrow$} & \textbf{FID$\downarrow$} \\
\midrule
U-ViT~\citep{bao2023all}            & 44M   & 11.3G & $1.0\times$ & 3.11 \\
\midrule
SparseDM (3:4)~\citep{wang2024sparsedm} & 11M   & 2.8G  & $4.0\times$ & 5.58 \\
Diff-Pruning~\citep{fang2023structural} & 14.5M & 3.7G  & $3.1\times$ & 4.63 \\
SparseDM (2:4)~\citep{wang2024sparsedm} & 22M   & 5.7G  & $2.0\times$ & \textbf{3.81} \\
\textbf{2ndM (Ours)}                & 14.5M & 3.7G  & $3.1\times$ & 4.05 \\
\midrule
\multicolumn{5}{c}{\textbf{CelebA ($64\times64$)}} \\
\textbf{Method} & \textbf{\# Param $\downarrow$} & \textbf{MACs$\downarrow$} & \textbf{Speed $\uparrow$} & \textbf{FID$\downarrow$} \\
\midrule
U-ViT~\citep{bao2023all}            & 44M   & 11.3G & $1.0\times$ & 2.87 \\
\midrule
SparseDM (3:4)~\citep{wang2024sparsedm} & 11M   & 2.8G  & $4.0\times$ & 5.05 \\
Diff-Pruning~\citep{fang2023structural} & 14.5M & 3.7G  & $3.1\times$ & 3.57 \\
SparseDM (2:4)~\citep{wang2024sparsedm} & 22M   & 5.7G  & $2.0\times$ & 3.52 \\
\textbf{2ndM (Ours)}                & 14.5M & 3.7G  & $3.1\times$ & \textbf{3.35} \\
\bottomrule
\end{tabular}
\vspace{-0.3cm}
\end{table}
On CIFAR-10, 2ndMatch achieves FID of $\textbf{4.05}$, improving over Diff-Pruning ($4.63$) while maintaining the same parameter count ($14.5$M) and MACs ($3.7$G). Compared to SparseDM ($3$:$4$), which achieves FID of $5.58$ with $4 \times$ speed, 2ndMatch offers a better balance between generation quality and efficiency. SparseDM ($2$:$4$) reduces the FID to $3.81$ but doubles the parameter count to $22$M. On CelebA, 2ndMatch attains FID of $\textbf{3.35}$, outperforming Diff-Pruning ($3.57$) and the two SparseDM variants ($5.05$ and $3.52$). Overall, 2ndMatch delivers consistent improvements in sample quality while maintaining competitive efficiency. These results further support that 2ndMatch is architecture agnostic and can be applied broadly across both convolutional and attention-based diffusion models.
\subsection{Ablation Study}
We conducted extensive ablation studies to analyze loss components and examine robustness and generality of 2ndMatch across pruning ratios and pruning methods.
\begin{table}[ht!]
\centering
\setlength{\tabcolsep}{3pt}
\renewcommand{\arraystretch}{1.05}
\caption{Loss components analysis on CIFAR-10. Each column adds one component cumulatively: NP → NP+KD → +1st JM → +2ndM (Ours).}
\vspace{-0.3cm}
\begin{tabular}{l|ccccc}
\toprule
 & \textbf{NP} & \textbf{NP+KD} & \textbf{+1st JM} & \textbf{+2ndM (Ours)} & \textbf{Dense} \\
\midrule
\textbf{FID} $\downarrow$ & 5.29 & 5.05 & 5.14 & \textbf{4.58} & 4.19 \\
\textbf{FTLE} & 0.413 & 0.418 & – & \textbf{0.408} & 0.405 \\
\bottomrule
\end{tabular}
\vspace{-0.6cm}
\label{tab:loss_analysis}
\end{table}
\paragraph{Loss Components Analysis.} We empirically analyzed the contribution of each loss component in Table~\ref{tab:loss_analysis}. Finetuning with NP + KD yields a clear improvement in FID, but the FTLE value remains essentially unchanged (0.418 vs.\ 0.413), indicating limited improvement in temporal sensitivity relative to the dense teacher (0.405). Adding first-order Jacobian matching (1st JM) does not further improve FID, which is consistent with the theoretical redundancy discussed earlier. In contrast, incorporating our second-order objective (\textbf{2ndM}) provides substantial benefits: it further reduces FID (4.58 vs.\ 5.05) and brings the FTLE closer to the dense model (0.408 vs.\ 0.405). These results demonstrate that second-order Jacobian matching enhances both generative quality and temporal sensitivity alignment.
\vspace{-0.4cm}
\paragraph{Robustness Across Pruning Ratios.}
Table~\ref{tab:pruning_ratios} compares the performance of Diff-Pruning (DP) and 2ndMatch on both U-Net and U-ViT architectures under varying pruning ratios (PR). Across all configurations, 2ndMatch consistently improves FID over Diff-Pruning, with larger gains observed at higher pruning ratios. For instance, on the U-Net model with 70\% pruning, 2ndMatch achieves FID of \textbf{7.10}, outperforming Diff-Pruning (9.33) by a significant margin. This trend holds for the U-ViT model as well, where 2ndMatch maintains superior fidelity even at aggressive pruning (e.g., \textbf{5.01} vs. 6.68 at 80\% pruning).
These results highlight that 2ndMatch effectively compensates for the representational loss caused by pruning, maintaining both perceptual quality and distributional alignment.
\begin{table}[t!]
\centering
\caption{Comparison of Diff-Pruning (DP) and 2ndM across pruning ratios (PR) on U-Net and U-ViT architectures.}
\vspace{-0.3cm}
\label{tab:pruning_ratios}
\setlength{\tabcolsep}{2pt}
\begin{tabular}{ccccccc}
\toprule
\multicolumn{7}{c}{\textbf{U-Net}}\\
\multirow{2}{*}{PR} & \multirow{2}{*}{\textbf{\# Param $\downarrow$}} & \multirow{2}{*}{\textbf{MACs$\downarrow$}} & \multirow{2}{*}{\textbf{Speed $\uparrow$}} & \multirow{2}{*}{\textbf{Imgs/s $\uparrow$}}  & \multicolumn{2}{c}{\textbf{FID} $\downarrow$} \\
 & & & & & DP & 2ndM \\
\midrule
dense & 35.7M & 6.1G & 1.0 x & 31.1 & \multicolumn{2}{c}{4.19} \\
0.56 & 14.3M & 2.7G & 1.46x & 45.5 &6.36 & \textbf{5.54} \\
0.70 & 8.6M  & 1.5G & 2.19x & 68.2 &9.33 & \textbf{7.10} \\
0.82 & 5.1M  & 1.1G & 2.58x & 80.2 &13.42 & \textbf{10.91} \\
\midrule
\multicolumn{6}{c}{\textbf{U-ViT}}\\
\multirow{2}{*}{PR} & \multirow{2}{*}{\textbf{\# Param $\downarrow$}} & \multirow{2}{*}{\textbf{MACs$\downarrow$}} & \multirow{2}{*}{\textbf{Speed $\uparrow$}} & \multirow{2}{*}{\textbf{Imgs/s $\uparrow$}}  & \multicolumn{2}{c}{\textbf{FID} $\downarrow$} \\
 & & & & & DP & 2ndM \\
\midrule
dense & 44M   & 11.3G & 1.00 & 0.54 & \multicolumn{2}{c}{3.11} \\
0.60  & 18.5M & 4.7G  & 1.80 & 0.97 & 3.83 & \textbf{3.75} \\
0.70  & 14.5M & 3.7G  & 2.07 & 1.12 & 4.63 & \textbf{4.05} \\
0.80  & 10.4M & 2.6G  & 2.44 & 1.32 & 6.68 & \textbf{5.01} \\
\bottomrule
\end{tabular}
\vspace{-0.5cm}
\end{table}
\vspace{-0.4cm}
\paragraph{Effect of Pruning Methods.} Table~\ref{tab:pruning_methods} further evaluates 2ndMatch across different pruning strategies on CIFAR-10 at 44\% pruning ratio.
Compared to standard finetuning and knowledge distillation (KD), incorporating 2ndMatch yields consistent FID improvements across all pruning methods. Even simple random pruning benefits from 2ndMatch approach (6.22 → 5.57), and more advanced and structured methods such as Diff-Pruning achieve their best performance with 2ndMatch (5.29 → 4.58).
This demonstrates that 2ndMatch is orthogonal to the underlying pruning criterion and can be seamlessly integrated with various sparsification techniques to enhance finetuning stability and image quality.
\begin{table}[ht]
\centering
\caption{FID scores on CIFAR-10 (44\% pruning ratio) for different pruning methods with and without inclusion of 2ndM. Lower is better.}
\vspace{-0.3cm}
\label{tab:pruning_methods}
\begin{tabular}{lccc}
\toprule
\textbf{Pruning Method} & \textbf{Baseline} & \textbf{KD} & \textbf{2ndM} \\
\midrule
Random Pruning     & 6.22 & 5.88 & \textbf{5.57} \\
LAMP Pruning       & 6.09 & 5.57 & \textbf{5.30} \\
Magnitude Pruning  & 5.74 & 5.53 & \textbf{5.25} \\
Taylor Pruning     & 5.71 & 5.39 & \textbf{5.17} \\
Diff-Pruning       & 5.29 & 5.05 & \textbf{4.58} \\
\bottomrule
\end{tabular}
\vspace{-0.3cm}
\end{table}
\vspace{-0.1cm}
\section{Conclusion}
In this work, we introduced \textbf{2ndMatch}, a general purpose finetuning framework for pruned diffusion models that is architecture and pruning-method agnostic. By jointly integrating noise prediction, knowledge distillation, and a second-order Jacobian matching objective inspired by Finite-Time Lyapunov Exponents (FTLE), 2ndMatch provides supervision on the model’s temporal sensitivity in addition to optimization of score prediction. Comprehensive experiments across U-Net and Transformer-based diffusion architectures—spanning diverse datasets and resolutions—demonstrate that \textbf{2ndMatch} consistently reduces the quality gap between pruned and dense models, offering an efficient and simple, yet powerful, route toward high-quality, resource-efficient generative diffusion systems.

\newpage
\section{Acknowledgment}
The authors acknowledge the partial support of HDR Institute: Accelerated AI Algorithms for
Data-Driven Discovery (A3D3) National Science Foundation grant PHY-2117997, and National Science Foundation grant EFRI-BRAID-2223495. The authors also
acknowledge the partial support by the Departments of Electrical Computer Engineering and Applied Mathematics. The authors are thankful to the eScience Center at the University of Washington.
{
    \small
    \bibliographystyle{ieeenat_fullname}
    \bibliography{ref}

@article{ho2020denoising,
  title={Denoising diffusion probabilistic models},
  author={Ho, Jonathan and Jain, Ajay and Abbeel, Pieter},
  journal={Advances in neural information processing systems},
  volume={33},
  pages={6840--6851},
  year={2020}
}

@article{saharia2022photorealistic,
  title={Photorealistic text-to-image diffusion models with deep language understanding},
  author={Saharia, Chitwan and Chan, William and Saxena, Saurabh and Li, Lala and Whang, Jay and Denton, Emily L and Ghasemipour, Kamyar and Gontijo Lopes, Raphael and Karagol Ayan, Burcu and Salimans, Tim and others},
  journal={Advances in Neural Information Processing Systems},
  volume={35},
  pages={36479--36494},
  year={2022}
}

@inproceedings{parmar2023zero,
  title={Zero-shot image-to-image translation},
  author={Parmar, Gaurav and Kumar Singh, Krishna and ZhangStructural Pruning for Diffusion Models, Richard and Li, Yijun and Lu, Jingwan and Zhu, Jun-Yan},
  booktitle={ACM SIGGRAPH 2023 Conference Proceedings},
  pages={1--11},
  year={2023}
}

@article{li2022srdiff,
  title={Srdiff: Single image super-resolution with diffusion probabilistic models},
  author={Li, Haoying and Yang, Yifan and Chang, Meng and Chen, Shiqi and Feng, Huajun and Xu, Zhihai and Li, Qi and Chen, Yueting},
  journal={Neurocomputing},
  volume={479},
  pages={47--59},
  year={2022},
  publisher={Elsevier}
}

@article{yang2023diffusion,
  title={Diffusion probabilistic modeling for video generation},
  author={Yang, Ruihan and Srivastava, Prakhar and Mandt, Stephan},
  journal={Entropy},
  volume={25},
  number={10},
  pages={1469},
  year={2023},
  publisher={MDPI}
}

@article{song2020denoising,
  title={Denoising diffusion implicit models},
  author={Song, Jiaming and Meng, Chenlin and Ermon, Stefano},
  journal={arXiv preprint arXiv:2010.02502},
  year={2020}
}

@article{song2020score,
  title={Score-based generative modeling through stochastic differential equations},
  author={Song, Yang and Sohl-Dickstein, Jascha and Kingma, Diederik P and Kumar, Abhishek and Ermon, Stefano and Poole, Ben},
  journal={arXiv preprint arXiv:2011.13456},
  year={2020}
}

@inproceedings{nichol2021improved,
  title={Improved denoising diffusion probabilistic models},
  author={Nichol, Alexander Quinn and Dhariwal, Prafulla},
  booktitle={International Conference on Machine Learning},
  pages={8162--8171},
  year={2021},
  organization={PMLR}
}

@inproceedings{shang2023post,
  title={Post-training quantization on diffusion models},
  author={Shang, Yuzhang and Yuan, Zhihang and Xie, Bin and Wu, Bingzhe and Yan, Yan},
  booktitle={Proceedings of the IEEE/CVF Conference on Computer Vision and Pattern Recognition},
  pages={1972--1981},
  year={2023}
}

@inproceedings{li2023q,
  title={Q-diffusion: Quantizing diffusion models},
  author={Li, Xiuyu and Liu, Yijiang and Lian, Long and Yang, Huanrui and Dong, Zhen and Kang, Daniel and Zhang, Shanghang and Keutzer, Kurt},
  booktitle={Proceedings of the IEEE/CVF International Conference on Computer Vision},
  pages={17535--17545},
  year={2023}
}

@article{fang2023structural,
  title={Structural Pruning for Diffusion Models},
  author={Fang, Gongfan and Ma, Xinyin and Wang, Xinchao},
  journal={arXiv preprint arXiv:2305.10924},
  year={2023}
}

@inproceedings{kim2024bk,
  title={Bk-sdm: A lightweight, fast, and cheap version of stable diffusion},
  author={Kim, Bo-Kyeong and Song, Hyoung-Kyu and Castells, Thibault and Choi, Shinkook},
  booktitle={European Conference on Computer Vision},
  pages={381--399},
  year={2024},
  organization={Springer}
}

@article{song2019generative,
  title={Generative modeling by estimating gradients of the data distribution},
  author={Song, Yang and Ermon, Stefano},
  journal={Advances in neural information processing systems},
  volume={32},
  year={2019}
}

@inproceedings{song2020sliced,
  title={Sliced score matching: A scalable approach to density and score estimation},
  author={Song, Yang and Garg, Sahaj and Shi, Jiaxin and Ermon, Stefano},
  booktitle={Uncertainty in Artificial Intelligence},
  pages={574--584},
  year={2020},
  organization={PMLR}
}

@inproceedings{srinivas2018knowledge,
  title={Knowledge transfer with jacobian matching},
  author={Srinivas, Suraj and Fleuret, Fran{\c{c}}ois},
  booktitle={International Conference on Machine Learning},
  pages={4723--4731},
  year={2018},
  organization={PMLR}
}

@article{hinton2015distilling,
  title={Distilling the knowledge in a neural network},
  author={Hinton, Geoffrey and Vinyals, Oriol and Dean, Jeff},
  journal={arXiv preprint arXiv:1503.02531},
  year={2015}
}

@article{ruelle1979ergodic,
  title={Ergodic theory of differentiable dynamical systems},
  author={Ruelle, David},
  journal={Publications Math{\'e}matiques de l'Institut des Hautes {\'E}tudes Scientifiques},
  volume={50},
  number={1},
  pages={27--58},
  year={1979},
  publisher={Springer}
}

@article{oseledets2008oseledets,
  title={Oseledets theorem},
  author={Oseledets, Valery},
  journal={Scholarpedia},
  volume={3},
  number={1},
  pages={1846},
  year={2008}
}

@article{shadden2005definition,
  title={Definition and properties of Lagrangian coherent structures from finite-time Lyapunov exponents in two-dimensional aperiodic flows},
  author={Shadden, Shawn C and Lekien, Francois and Marsden, Jerrold E},
  journal={Physica D: Nonlinear Phenomena},
  volume={212},
  number={3-4},
  pages={271--304},
  year={2005},
  publisher={Elsevier}
}

@article{brunton2010fast,
  title={Fast computation of finite-time Lyapunov exponent fields for unsteady flows},
  author={Brunton, Steven L and Rowley, Clarence W},
  journal={Chaos: An Interdisciplinary Journal of Nonlinear Science},
  volume={20},
  number={1},
  year={2010},
  publisher={AIP Publishing}
}

@article{lipinski2010ridge,
  title={A ridge tracking algorithm and error estimate for efficient computation of Lagrangian coherent structures},
  author={Lipinski, Doug and Mohseni, Kamran},
  journal={Chaos: An Interdisciplinary Journal of Nonlinear Science},
  volume={20},
  number={1},
  pages={017504},
  year={2010},
  publisher={American Institute of Physics}
}

@article{haller2015lagrangian,
  title={Lagrangian coherent structures},
  author={Haller, George},
  journal={Annual review of fluid mechanics},
  volume={47},
  pages={137--162},
  year={2015},
  publisher={Annual Reviews}
}

@article{vogt2024lyapunov,
  title={Lyapunov-guided representation of recurrent neural network performance},
  author={Vogt, Ryan and Zheng, Yang and Shlizerman, Eli},
  journal={Neural Computing and Applications},
  pages={1--16},
  year={2024},
  publisher={Springer}
}

@article{lu2022dpm,
  title={Dpm-solver: A fast ode solver for diffusion probabilistic model sampling in around 10 steps},
  author={Lu, Cheng and Zhou, Yuhao and Bao, Fan and Chen, Jianfei and Li, Chongxuan and Zhu, Jun},
  journal={Advances in Neural Information Processing Systems},
  volume={35},
  pages={5775--5787},
  year={2022}
}

@article{zhang2022fast,
  title={Fast sampling of diffusion models with exponential integrator},
  author={Zhang, Qinsheng and Chen, Yongxin},
  journal={arXiv preprint arXiv:2204.13902},
  year={2022}
}

@article{watson2021learning,
  title={Learning to efficiently sample from diffusion probabilistic models},
  author={Watson, Daniel and Ho, Jonathan and Norouzi, Mohammad and Chan, William},
  journal={arXiv preprint arXiv:2106.03802},
  year={2021}
}

@article{salimans2022progressive,
  title={Progressive distillation for fast sampling of diffusion models},
  author={Salimans, Tim and Ho, Jonathan},
  journal={arXiv preprint arXiv:2202.00512},
  year={2022}
}

@article{luhman2021knowledge,
  title={Knowledge distillation in iterative generative models for improved sampling speed},
  author={Luhman, Eric and Luhman, Troy},
  journal={arXiv preprint arXiv:2101.02388},
  year={2021}
}

@article{czarnecki2017sobolev,
  title={Sobolev training for neural networks},
  author={Czarnecki, Wojciech M and Osindero, Simon and Jaderberg, Max and Swirszcz, Grzegorz and Pascanu, Razvan},
  journal={Advances in neural information processing systems},
  volume={30},
  year={2017}
}

@article{hyvarinen2005estimation,
  title={Estimation of non-normalized statistical models by score matching.},
  author={Hyv{\"a}rinen, Aapo and Dayan, Peter},
  journal={Journal of Machine Learning Research},
  volume={6},
  number={4},
  year={2005}
}

@inproceedings{rombach2022high,
  title={High-resolution image synthesis with latent diffusion models},
  author={Rombach, Robin and Blattmann, Andreas and Lorenz, Dominik and Esser, Patrick and Ommer, Bj{\"o}rn},
  booktitle={Proceedings of the IEEE/CVF conference on computer vision and pattern recognition},
  pages={10684--10695},
  year={2022}
}

@inproceedings{peebles2023scalable,
  title={Scalable diffusion models with transformers},
  author={Peebles, William and Xie, Saining},
  booktitle={Proceedings of the IEEE/CVF international conference on computer vision},
  pages={4195--4205},
  year={2023}
}

@inproceedings{castells2024ld,
  title={Ld-pruner: Efficient pruning of latent diffusion models using task-agnostic insights},
  author={Castells, Thibault and Song, Hyoung-Kyu and Kim, Bo-Kyeong and Choi, Shinkook},
  booktitle={Proceedings of the IEEE/CVF Conference on Computer Vision and Pattern Recognition},
  pages={821--830},
  year={2024}
}

@article{zhang2024laptop,
  title={Laptop-diff: Layer pruning and normalized distillation for compressing diffusion models},
  author={Zhang, Dingkun and Li, Sijia and Chen, Chen and Xie, Qingsong and Lu, Haonan},
  journal={arXiv preprint arXiv:2404.11098},
  year={2024}
}

@article{zhang2024effortless,
  title={Effortless Efficiency: Low-Cost Pruning of Diffusion Models},
  author={Zhang, Yang and Jin, Er and Dong, Yanfei and Khakzar, Ashkan and Torr, Philip and Stegmaier, Johannes and Kawaguchi, Kenji},
  journal={arXiv preprint arXiv:2412.02852},
  year={2024}
}

@article{zhu2024dip,
  title={Dip-go: A diffusion pruner via few-step gradient optimization},
  author={Zhu, Haowei and Tang, Dehua and Liu, Ji and Lu, Mingjie and Zheng, Jintu and Peng, Jinzhang and Li, Dong and Wang, Yu and Jiang, Fan and Tian, Lu and others},
  journal={Advances in Neural Information Processing Systems},
  volume={37},
  pages={92581--92604},
  year={2024}
}

@article{karras2022elucidating,
  title={Elucidating the design space of diffusion-based generative models},
  author={Karras, Tero and Aittala, Miika and Aila, Timo and Laine, Samuli},
  journal={Advances in neural information processing systems},
  volume={35},
  pages={26565--26577},
  year={2022}
}

@inproceedings{liu2023oms,
  title={Oms-dpm: Optimizing the model schedule for diffusion probabilistic models},
  author={Liu, Enshu and Ning, Xuefei and Lin, Zinan and Yang, Huazhong and Wang, Yu},
  booktitle={International Conference on Machine Learning},
  pages={21915--21936},
  year={2023},
  organization={PMLR}
}

@article{pan2024t,
  title={T-stitch: Accelerating sampling in pre-trained diffusion models with trajectory stitching},
  author={Pan, Zizheng and Zhuang, Bohan and Huang, De-An and Nie, Weili and Yu, Zhiding and Xiao, Chaowei and Cai, Jianfei and Anandkumar, Anima},
  journal={arXiv preprint arXiv:2402.14167},
  year={2024}
}

@inproceedings{wimbauer2024cache,
  title={Cache me if you can: Accelerating diffusion models through block caching},
  author={Wimbauer, Felix and Wu, Bichen and Schoenfeld, Edgar and Dai, Xiaoliang and Hou, Ji and He, Zijian and Sanakoyeu, Artsiom and Zhang, Peizhao and Tsai, Sam and Kohler, Jonas and others},
  booktitle={Proceedings of the IEEE/CVF Conference on Computer Vision and Pattern Recognition},
  pages={6211--6220},
  year={2024}
}

@inproceedings{ma2024deepcache,
  title={Deepcache: Accelerating diffusion models for free},
  author={Ma, Xinyin and Fang, Gongfan and Wang, Xinchao},
  booktitle={Proceedings of the IEEE/CVF conference on computer vision and pattern recognition},
  pages={15762--15772},
  year={2024}
}

@inproceedings{agarwal2024approximate,
  title={Approximate caching for efficiently serving $\{$Text-to-Image$\}$ diffusion models},
  author={Agarwal, Shubham and Mitra, Subrata and Chakraborty, Sarthak and Karanam, Srikrishna and Mukherjee, Koyel and Saini, Shiv Kumar},
  booktitle={21st USENIX Symposium on Networked Systems Design and Implementation (NSDI 24)},
  pages={1173--1189},
  year={2024}
}

@article{wang2024sparsedm,
  title={Sparsedm: Toward sparse efficient diffusion models},
  author={Wang, Kafeng and Chen, Jianfei and Li, He and Mi, Zhenpeng and Zhu, Jun},
  journal={arXiv preprint arXiv:2404.10445},
  year={2024}
}

@inproceedings{bao2023all,
  title={All are worth words: A vit backbone for diffusion models},
  author={Bao, Fan and Nie, Shen and Xue, Kaiwen and Cao, Yue and Li, Chongxuan and Su, Hang and Zhu, Jun},
  booktitle={Proceedings of the IEEE/CVF conference on computer vision and pattern recognition},
  pages={22669--22679},
  year={2023}
}

@article{zhang2024accelerating,
  title={Accelerating Diffusion Models with One-to-Many Knowledge Distillation},
  author={Zhang, Linfeng and Ma, Kaisheng},
  journal={arXiv preprint arXiv:2410.04191},
  year={2024}
}

@article{zheng2025hyperpruning,
  title={Hyperpruning: Efficient Search through Pruned Variants of Recurrent Neural Networks Leveraging Lyapunov Spectrum},
  author={Zheng, Caleb and Shlizerman, Eli},
  journal={arXiv preprint arXiv:2506.07975},
  year={2025}
}

@article{vincent2011connection,
  title={A connection between score matching and denoising autoencoders},
  author={Vincent, Pascal},
  journal={Neural computation},
  volume={23},
  number={7},
  pages={1661--1674},
  year={2011},
  publisher={MIT Press}
}

@misc{krizhevsky2009learning,
  title={Learning multiple layers of features from tiny images.(2009)},
  author={Krizhevsky, Alex and Hinton, Geoffrey and others},
  year={2009}
}

@article{yu2015lsun,
  title={Lsun: Construction of a large-scale image dataset using deep learning with humans in the loop},
  author={Yu, Fisher and Seff, Ari and Zhang, Yinda and Song, Shuran and Funkhouser, Thomas and Xiao, Jianxiong},
  journal={arXiv preprint arXiv:1506.03365},
  year={2015}
}

@inproceedings{liu2015deep,
  title={Deep learning face attributes in the wild},
  author={Liu, Ziwei and Luo, Ping and Wang, Xiaogang and Tang, Xiaoou},
  booktitle={Proceedings of the IEEE international conference on computer vision},
  pages={3730--3738},
  year={2015}
}

@article{heusel2017gans,
  title={Gans trained by a two time-scale update rule converge to a local nash equilibrium},
  author={Heusel, Martin and Ramsauer, Hubert and Unterthiner, Thomas and Nessler, Bernhard and Hochreiter, Sepp},
  journal={Advances in neural information processing systems},
  volume={30},
  year={2017}
}

@article{salimans2016improved,
  title={Improved techniques for training gans},
  author={Salimans, Tim and Goodfellow, Ian and Zaremba, Wojciech and Cheung, Vicki and Radford, Alec and Chen, Xi},
  journal={Advances in neural information processing systems},
  volume={29},
  year={2016}
}

@article{cai2025shortcutting,
  title={Shortcutting Pre-trained Flow Matching Diffusion Models is Almost Free Lunch},
  author={Cai, Xu and Wu, Yang and Chen, Qianli and Wu, Haoran and Xiang, Lichuan and Wen, Hongkai},
  journal={arXiv preprint arXiv:2510.17858},
  year={2025}
}

@article{zheng2024diffusion,
  title={Diffusion models are innate one-step generators},
  author={Zheng, Bowen and Yang, Tianming},
  journal={arXiv preprint arXiv:2405.20750},
  year={2024}
}

@article{ganjdanesh2024not,
  title={Not all prompts are made equal: Prompt-based pruning of text-to-image diffusion models},
  author={Ganjdanesh, Alireza and Shirkavand, Reza and Gao, Shangqian and Huang, Heng},
  journal={arXiv preprint arXiv:2406.12042},
  year={2024}
}

@inproceedings{shirkavand2025efficient,
  title={Efficient fine-tuning and concept suppression for pruned diffusion models},
  author={Shirkavand, Reza and Yu, Peiran and Gao, Shangqian and Somepalli, Gowthami and Goldstein, Tom and Huang, Heng},
  booktitle={Proceedings of the Computer Vision and Pattern Recognition Conference},
  pages={18619--18629},
  year={2025}
}

@article{nash2021generating,
  title={Generating images with sparse representations},
  author={Nash, Charlie and Menick, Jacob and Dieleman, Sander and Battaglia, Peter W},
  journal={arXiv preprint arXiv:2103.03841},
  year={2021}
}

@inproceedings{hessel2021clipscore,
  title={Clipscore: A reference-free evaluation metric for image captioning},
  author={Hessel, Jack and Holtzman, Ari and Forbes, Maxwell and Le Bras, Ronan and Choi, Yejin},
  booktitle={Proceedings of the 2021 conference on empirical methods in natural language processing},
  pages={7514--7528},
  year={2021}
}

@article{kynkaanniemi2019improved,
  title={Improved precision and recall metric for assessing generative models},
  author={Kynk{\"a}{\"a}nniemi, Tuomas and Karras, Tero and Laine, Samuli and Lehtinen, Jaakko and Aila, Timo},
  journal={Advances in neural information processing systems},
  volume={32},
  year={2019}
}

@inproceedings{deng2009imagenet,
  title={Imagenet: A large-scale hierarchical image database},
  author={Deng, Jia and Dong, Wei and Socher, Richard and Li, Li-Jia and Li, Kai and Fei-Fei, Li},
  booktitle={2009 IEEE conference on computer vision and pattern recognition},
  pages={248--255},
  year={2009},
  organization={Ieee}
}

@inproceedings{lin2014microsoft,
  title={Microsoft coco: Common objects in context},
  author={Lin, Tsung-Yi and Maire, Michael and Belongie, Serge and Hays, James and Perona, Pietro and Ramanan, Deva and Doll{\'a}r, Piotr and Zitnick, C Lawrence},
  booktitle={European conference on computer vision},
  pages={740--755},
  year={2014},
  organization={Springer}
}

@inproceedings{fang2025tinyfusion,
  title={Tinyfusion: Diffusion transformers learned shallow},
  author={Fang, Gongfan and Li, Kunjun and Ma, Xinyin and Wang, Xinchao},
  booktitle={Proceedings of the Computer Vision and Pattern Recognition Conference},
  pages={18144--18154},
  year={2025}
}

@inproceedings{ma2025diffusion,
  title={Diffusion Model is Effectively Its Own Teacher},
  author={Ma, Xinyin and Yu, Runpeng and Liu, Songhua and Fang, Gongfan and Wang, Xinchao},
  booktitle={Proceedings of the Computer Vision and Pattern Recognition Conference},
  pages={12901--12911},
  year={2025}
}

@inproceedings{zhao2025pioneering,
  title={Pioneering 4-Bit FP Quantization for Diffusion Models: Mixup-Sign Quantization and Timestep-Aware Fine-Tuning},
  author={Zhao, Maosen and Chen, Pengtao and Yu, Chong and Wen, Yan and Tan, Xudong and Chen, Tao},
  booktitle={Proceedings of the Computer Vision and Pattern Recognition Conference},
  pages={18134--18143},
  year={2025}
}

@inproceedings{chen2025q,
  title={Q-dit: Accurate post-training quantization for diffusion transformers},
  author={Chen, Lei and Meng, Yuan and Tang, Chen and Ma, Xinzhu and Jiang, Jingyan and Wang, Xin and Wang, Zhi and Zhu, Wenwu},
  booktitle={Proceedings of the Computer Vision and Pattern Recognition Conference},
  pages={28306--28315},
  year={2025}
}

@inproceedings{ye2025schedule,
  title={Schedule on the fly: Diffusion time prediction for faster and better image generation},
  author={Ye, Zilyu and Chen, Zhiyang and Li, Tiancheng and Huang, Zemin and Luo, Weijian and Qi, Guo-Jun},
  booktitle={Proceedings of the Computer Vision and Pattern Recognition Conference},
  pages={23412--23422},
  year={2025}
}

@inproceedings{ding2025rass,
  title={RaSS: Improving Denoising Diffusion Samplers with Reinforced Active Sampling Scheduler},
  author={Ding, Xin and Yu, Lei and Li, Xin and Tu, Zhijun and Chen, Hanting and Hu, Jie and Chen, Zhibo},
  booktitle={Proceedings of the Computer Vision and Pattern Recognition Conference},
  pages={12923--12933},
  year={2025}
}

@inproceedings{liu2025timestep,
  title={Timestep Embedding Tells: It's Time to Cache for Video Diffusion Model},
  author={Liu, Feng and Zhang, Shiwei and Wang, Xiaofeng and Wei, Yujie and Qiu, Haonan and Zhao, Yuzhong and Zhang, Yingya and Ye, Qixiang and Wan, Fang},
  booktitle={Proceedings of the Computer Vision and Pattern Recognition Conference},
  pages={7353--7363},
  year={2025}
}

@inproceedings{meng2023distillation,
  title={On distillation of guided diffusion models},
  author={Meng, Chenlin and Rombach, Robin and Gao, Ruiqi and Kingma, Diederik and Ermon, Stefano and Ho, Jonathan and Salimans, Tim},
  booktitle={Proceedings of the IEEE/CVF conference on computer vision and pattern recognition},
  pages={14297--14306},
  year={2023}
}

@inproceedings{wang2025videoscene,
  title={Videoscene: Distilling video diffusion model to generate 3d scenes in one step},
  author={Wang, Hanyang and Liu, Fangfu and Chi, Jiawei and Duan, Yueqi},
  booktitle={2025 IEEE/CVF Conference on Computer Vision and Pattern Recognition (CVPR)},
  pages={16475--16485},
  year={2025},
  organization={IEEE}
}
}


\end{document}